\documentclass[10pt ,a4]{article}

\usepackage{fullpage}
\usepackage{setspace}
\usepackage{parskip}
\usepackage{titlesec}
\usepackage{xcolor}
\usepackage{lineno}

\PassOptionsToPackage{hyphens}{url}
\usepackage[colorlinks = true,
            linkcolor = blue,
            urlcolor  = blue,
            citecolor = blue,
            anchorcolor = blue]{hyperref}
\usepackage{etoolbox}
\makeatletter
\patchcmd\@combinedblfloats{\box\@outputbox}{\unvbox\@outputbox}{}{%
  \errmessage{\noexpand\@combinedblfloats could not be patched}%
}%
\makeatother

\usepackage[round]{natbib}
\let\cite\citep

\renewenvironment{abstract}
  {{\bfseries\noindent{\abstractname}\par\nobreak}\footnotesize}
  {\bigskip}

\titlespacing{\section}{0pt}{*3}{*1}
\titlespacing{\subsection}{0pt}{*2}{*0.5}
\titlespacing{\subsubsection}{0pt}{*1.5}{0pt}

\usepackage{authblk}

\usepackage{graphicx}
\usepackage[space]{grffile}
\usepackage{latexsym}
\usepackage{textcomp}
\usepackage{longtable}
\usepackage{tabulary}
\usepackage{booktabs,array,multirow}
\usepackage{amsfonts,amsmath,amssymb}
\providecommand\citet{\cite}
\providecommand\citep{\cite}

\newif\iflatexml\latexmlfalse

\AtBeginDocument{\DeclareGraphicsExtensions{.pdf,.PDF,.eps,.EPS,.png,.PNG,.tif,.TIF,.jpg,.JPG,.jpeg,.JPEG}}

\usepackage[utf8]{inputenc}
\usepackage[ngerman,english]{babel}

\usepackage[printfigures]{figcaps}

\begin{document}

\title{Disrupted core-periphery structure of multimodal brain networks in
Alzheimer's Disease}

\author[1,2]{Jeremy Guillon}%
\author[3]{Mario Chavez}%
\author[10,3,2]{Federico Battiston}%
\author[4]{Yohan Attal}%
\author[5,6,7]{Valentina La Corte}%
\author[1]{Michel Thiebaut de Schotten}%
\author[8]{Bruno Dubois}%
\author[9]{Denis Schwartz}%
\author[2,1]{Olivier Colliot}%
\author[2,1,*]{Fabrizio De Vico Fallani}%
\affil[1]{Institut du Cerveau et de la Moelle Epiniere, ICM, Inserm, U 1127, CNRS, UMR 7225, Sorbonne Universite, F-75013, Paris, France}
\affil[2]{Inria Paris, Aramis project-team, F-75013, Paris, France}
\affil[3]{CNRS, UMR 7225, F-75013, Paris, France}
\affil[4]{MyBrain Technologies, Paris, France}
\affil[5]{Department of Neurology, Institut de la Memoire et de la Maladie d’Alzheimer - IM2A, Paris, France}%
\affil[6]{INSERM UMR 894, Center of Psychiatry and Neurosciences, Memory and Cognition Laboratory, Paris, France}%
\affil[7]{Institute of Psychology, University Paris Descartes, Sorbonne Paris Cite, France}%
\affil[8]{Institut de la Mémoire et de la Maladie d’Alzheimer - IM2A, AP-HP, Sorbonne Université, Paris, France}%
\affil[9]{Institut du Cerveau et de la Moelle Epiniere, ICM, Inserm U 1127, CNRS UMR 7225, Sorbonne Universite, Ecole Normale Superieure, ENS, Centre MEG-EEG, F-75013, Paris, France}
\affil[10]{Department of Network and Data Science, Central European University, Budapest 1051, Hungary}
\affil[*]{Corresponding author}

\vspace{-1em}

  \date{\today}

\begingroup
\let\center\flushleft
\let\endcenter\endflushleft
\maketitle
\endgroup


\selectlanguage{english}
\begin{abstract}

In Alzheimer’s disease (AD), the progressive atrophy leads to aberrant network reconfigurations both at structural and functional levels. In such network reorganization, the core and peripheral nodes appear to be crucial for the prediction of clinical outcome due to their ability to influence large-scale functional integration. However, the role of the different types of brain connectivity in such prediction still remains unclear. Using a multiplex network approach we integrated information from DWI, fMRI and MEG brain connectivity to extract an enriched description of the core-periphery structure in a group of AD patients and age- matched controls. Globally, the regional coreness - i.e., the probability of a region to be in the multiplex core - significantly decreased in AD patients as a result of the randomization process initiated by the neurodegeneration. Locally, the most impacted areas were in the core of the network - including temporal, parietal and occipital areas - while we reported compensatory increments for the peripheral regions in the sensorimotor system. Furthermore, these network changes significantly predicted the cognitive and memory impairment of patients. Taken together these results indicate that a more accurate description of neurodegenerative diseases can be obtained from the multimodal integration of neuroimaging-derived network data.

\par\null\par\null%
\end{abstract}%

\section*{Introduction}\label{introduction}

The brain is a complex network where differently specialized areas are
anatomically and functionally connected. Because of such interconnected
structure, focal damages can affect the rest of the network through the
interruption of communication pathways. Indeed, many neurological
disorders affecting language, motor and sensory abilities are often due
to a disconnection syndrome caused by the anatomical connectivity
breakdown between the relevant brain areas \cite{SCHMAHMANN_2008,geschwind_disconnexion_1965}. In the
case of neurodegenerative diseases, the disconnection hypothesis is
supported by a progressive death of neurons and synapses
that induce gross atrophy. Empirical evidence has shown that Alzheimer's
disease (AD) patients with severe motor and cognitive impairments
exhibited anatomical disconnections among regions between cerebral
hemispheres that resemble those observed in split-brain subjects
\cite{Delbeuck_2007, Lakmache_1998}. In Parkinson's disease (PD) intrahemispheric
dissociations between subcortical and cortical structures have been
linked to disturbances in cognition, perception, emotion, and sleep
\cite{Cronin_Golomb_2010}. In addition, functional connectivity alterations
within and between hemispheres have been reported in both AD
\cite{Blinowska_2017,Sankari_2010,Adler_2003,Babiloni_2009} and PD \cite{Luo_2015,Olde_Dubbelink_2013} suggesting their potential
role in the early diagnosis.

Altogether, these findings suggest that neurodegenerative diseases must
be considered as a network problem. Recent approaches based on network
theory have greatly advanced our understanding of the
connection mechanisms characterizing brain diseases \cite{Stam_2014}.
Among others, decreased efficiency, modularity and hub centrality have
been largely reported in neurodegeneration and associated with the stage
of disease. Increasing evidence suggests that the core-periphery
structure of the human connectome - supporting global integration of
information among distant areas - is highly affected by the AD process
and that resulting changes might be effective predictors of cognitive
declines. On one hand, brain areas forming the core of the network -
i.e.~central and mutually connected nodes - have
been reported to be preferentially attacked by AD \cite{Yan_2018}. On
the other hand, brain regions forming the periphery of the network -
i.e.~nodes that are only weakly connected to the other units in the network -
appear to be crucial for the degeneration \cite{Daianu_2015}. While
these results refer to structural brain connectivity, the relative
contribution of functional brain connectivity into the network
core-periphery changes remain poorly understood.

Based on the aforementioned empirical and theoretical grounds, we
hypotesize that neurodegeneration would affect the core-periphery
structure of the brain network at both anatomical and functional levels.
More specifically, we expected that the extraction of the core-periphery
organization by integrating information from multimodal brain networks
would give more accurate predictor of AD and cognitive impairment.
Finally, based on the evidence that hubs are the most attacked nodes in
many neurological diseases and psychiatric disorders \cite{van_den_Heuvel_2013a},
we hypothesize that the core brain regions would be mostly impacted by
the AD atrophy process.

To test these predictions, we considered multiple brain networks derived
from DWI, fMRI and MEG data recorded in a group of AD patients and
age-matched healthy controls (HC). Cognitive impairments in AD patients
were described using multidomain behavioral measurements. We extracted
the multimodal core-periphery structure of the brain networks through a
multiplex network approach, where all the available information is kept
at different connectivity layers. We evaluated how AD impacted the
multiplex core-periphery organization and we tested the correlation of
the regional coreness with the cognitive and memory impairment of
patients. See Material and methods for more details on the experimental
design and methods of analysis.\selectlanguage{english}
\begin{figure*}[h!]
\begin{center}
\includegraphics[width=\textwidth]{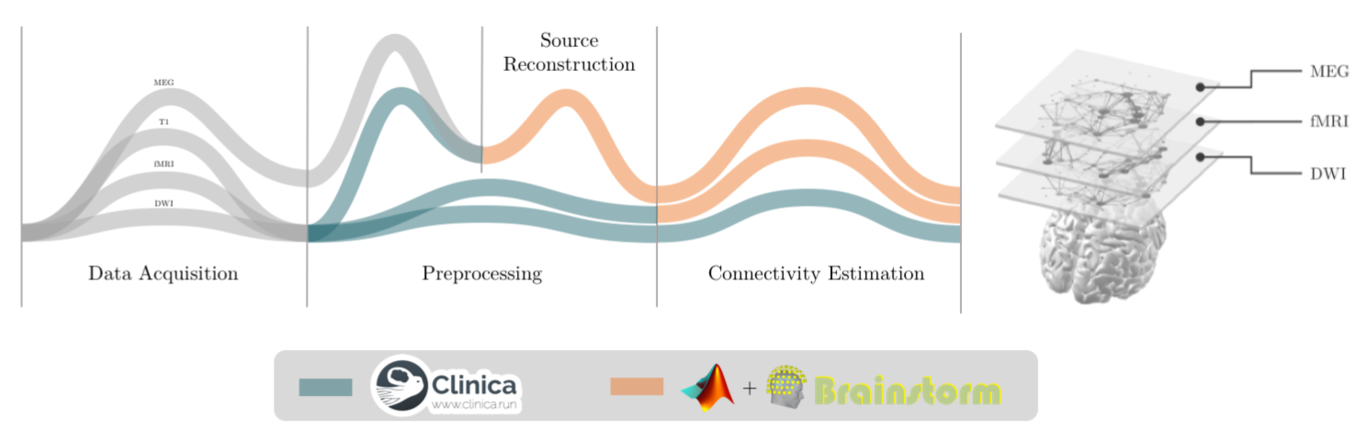}
\caption{{\textbf{Multiplex brain network construction.}~Different neuroimaging
data are collected and preprocessed separately. We used the Desikan
cortical atlas parcellation \protect\cite{Desikan_2006} to infer connectivity
networks from DWI, fMRI and MEG source-reconstructed data. The color of
the line indicates the software that has been used in each step of the
pipeline. We spatially aligned all the estimated brain networks to
construct the multiplex brain network.
{\label{349594}}%
}}
\end{center}
\end{figure*}

\section*{Results}\label{results}

\subsection*{Multimodal core of brain
networks}\label{multimodal-core-of-brain-networks}

We integrated multimodal information by constructing nine-layer
multiplex brain networks containing DWI, fMRI and MEG connectivity
between 68 cortical regions of interest (ROIs) (Material and methods).
To estimate the likelihood of each ROI \(i\) to be in the
multiplex core we computed its \emph{coreness} \(\mathcal{C}_i\) by
counting how many times it was in the multiplex core across different
density thresholds \cite{Battiston_2018}. At each threshold, the multiplex
core-periphery structure was obtained by linearly combining the node
strength of all the layers through a vector parameter \(c\)
(Material and methods).

\begin{figure*}[h!]
\begin{center}
\includegraphics[width=\textwidth]{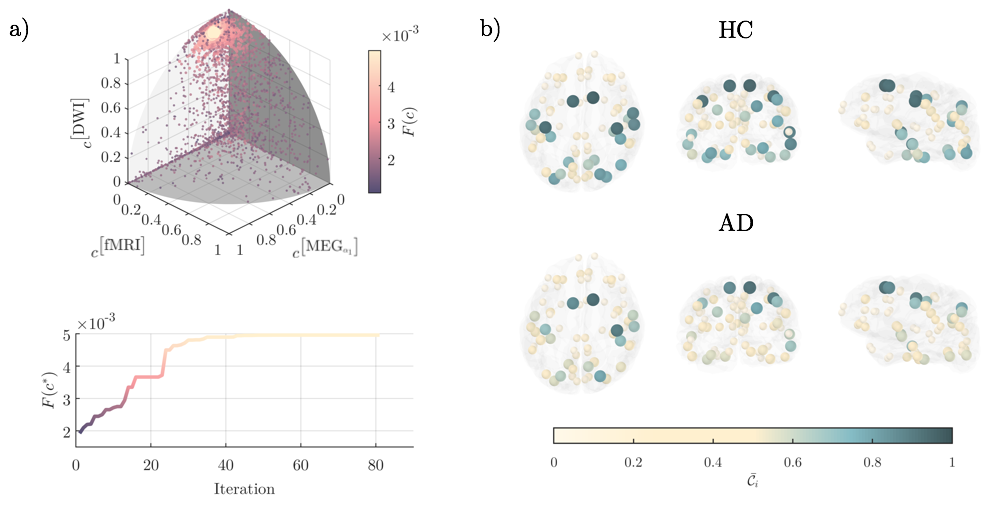}
\caption{{\textbf{Regional coreness of the multiplex brain networks}.
Panel~\textbf{a)}, shows the results of the particle~swarm optimization
(PSO) used to find the best layer coefficients vector~\(c\)
that maximizes the Fisher score~\(F(c)\) between AD and HC
subjects. In the upper plot, each dot represents the position of a
particle~at a given iteration in the original 9-dimension coreness
contribution coefficient vector space. The color of the dots code for
the corresponding Fisher score. Results were projected over the three
main network layers for the sake of illustration The other non-shown
components were rapidly zeroing-out during the 81 iterations needed to
converge to the optimum as shown in the bottom plot. Panel~\textbf{b)}
shows the\textbf{~}corresponding\textbf{~}average coreness~for~the
healthy control (HC) population and for the Alzheimer's disease (AD)
group. The position of the nodes identifies the barycentre of each ROI
in the cortical surface here represented in transparent grey; the color
of each node codes for the average coreness~\(\bar{\mathcal{C}}_i\)~.
{\label{661774}}%
}}
\end{center}
\end{figure*}

Because we do not know a-priori the best combination, we derived the
optimal \(c^{*}\) by using a data-driven approach that
efficiently explores the parameter space to maximize the difference
between AD and HC regional coreness. Specifically, we used the particles
swarm optimization algorithm (PSO) to maximize the Fisher's criterion
\(F(c)\) (Material and methods). Results show that the optimal
\(c^{*}\) components are found to be highly heterogenous and
that the DWI layer, as well as MEG-alpha1 and fMRI layers, are the main contributors
to separate the AD and HC group (Table
\ref{corenesscoefftable}, Figure \ref{661774}a).\selectlanguage{english}
\begin{table}[]
\caption{{Vector of the optimal layers weight for the coreness computation.}}
\label{corenesscoefftable}
\begin{tabular}{r|c}
Layer $m$ & $c^{*[m]}$ \\ \hline\hline
MEG$_{\delta}$ & 0.000 \\
MEG$_{\theta}$ & 0.001 \\
MEG$_{\alpha_1}$ & 0.258 \\
MEG$_{\alpha_2}$ & 0.000 \\
MEG$_{\beta_1}$ & 0.000 \\
MEG$_{\beta_2}$ & 0.002 \\
MEG$_{\gamma}$ & 0.000 \\
fMRI & 0.104 \\
DWI & 0.961
\end{tabular}
\end{table}

In the HC group, the multiplex core tended to include large portions of
temporal, superior parietal and occipital cortices, and to a minor
extent central and superior frontal regions (Figure \ref{661774}b). On
average AD patients exhibited a loss of coreness with respect to HC
particularly in the temporal, superior parietal and occipital cortices.
These regions were already known to form the core of multiplex brain
networks derived from DTI and fMRI data \cite{Battiston_2018}.\selectlanguage{english}

\subsection*{Reorganization of core-periphery structure in
AD}\label{reorganization-of-core-periphery-structure-in-ad}

\begin{figure*}[h!]
\begin{center}
\includegraphics[width=\textwidth]{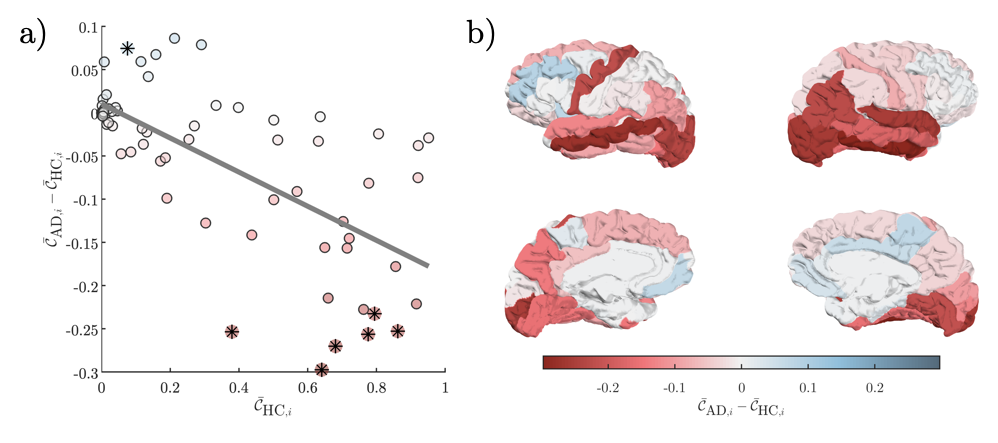}
\caption{{\textbf{Differences in regional coreness between AD and HC
subjects.}~Panel~\textbf{a)} shows the between-group difference of
coreness~\(\bar{\mathcal{C}}_{\textrm{AD},i}  - \bar{\mathcal{C}}_{\textrm{HC},i} \) as a function of the healthy population's
coreness~\(\bar{\mathcal{C}}_{\textrm{HC},i} \); the slope of the regression line in grey
measures the coreness disruption index~\(\kappa=-0.20\) . The color of
the circles~code for the difference between average coreness in the AD
and HC group; stars point out the ROIs for which we reported a
significant difference (\(p<0.025\)), Supplementary Table 2).
Panel~\textbf{b)}~illustrates the values of the between-group coreness
difference over the Desikan cortical atlas. Color code is the same in
Panel~\textbf{a)}.~
{\label{482763}}%
}}
\end{center}
\end{figure*}

To quantify the observed network changes we defined the coreness
disruption index \(\kappa\) as the slope of the line obtained by
regressing the difference between the average coreness (at each ROI and across subjects) of the two groups with the average coreness of the healthy one \cite{Termenon_2016}
(Material and methods). We found a significant negative
\(\kappa\) value indicating that AD preferentially attack ROIs
with a high coreness (\(\kappa=-0.20,p=2.45\mathrm{e}{-10}\)). This result was also
consistent at the individual level when we extracted the coreness
disruption index in each patient (Supplementary Table 1). In particular,
by statistically comparing the average coreness of the two groups, we reported
a significant decrease of coreness in core regions, such as temporal, parietal and
occipital cortices as well as a significant increase of coreness in the right
paracentral area which are instead more peripheral, (\(p<0.025\),
Figure \ref{482763}a,b, Supplementary Table 2).\selectlanguage{english}

\begin{figure}[h!]
\begin{center}
\includegraphics[width=250pt]{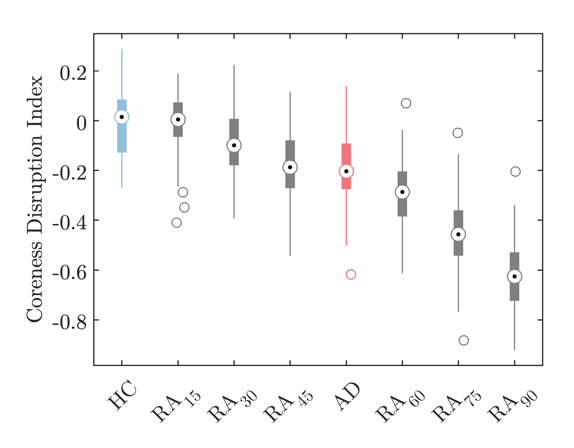}
\caption{{\textbf{Coreness disruption index as a function~of network
randomization}. Boxplots show the values of coreness disruption index
(\(\kappa\)) obtained by randomly~rewiring an increasing
percentage of links in the multiplex brain networks of the HC group (see
Materials and methods for more details). For example,~\(\text{RA}_{15}\)
means that 15\% of the links were reshuffled randomly in each layer. The
blue and red boxplots illustrate respectively the~\(\kappa\)
values for the HC and AD group. The circles in the boxes show the
median; the bottom and top edges of the boxes denote the 25th and
75th~percentile, respectively. Whiskers connect the most extreme points
not considered outliers,~and outliers are plotted individually as
circles.
{\label{722240}}%
}}
\end{center}
\end{figure}

Based on the hypothesis that AD is a disconnection syndrome \cite{geschwind_disconnexion_1965,delbeuck_alzheimers_2003} leading to
disorganized network configurations \cite{sanz-arigita_loss_2010}, we next generated a series of
synthetic multiplex networks starting from the ones observed in the HC
group and then progressively randomizing the same amount of links in
each layer (Materials and methods). As expected the coreness disruption
index decreased with the percentage of links that was randomly rewired.
Notably, we could obtain the same significantly impacted
\(\kappa\) values observed in the multiplex brain networks of
the AD group (\(p=2.77\mathrm{e}{-4}\)) by rewiring between 45\% and 60\% of
the links in the HC multiplex brain networks (Figure \ref{722240}).
Altogether, these findings indicate that the AD is associated with a
pervasive random reconfiguration of structural and functional connectivity that primarly affects the nodes of the
multiplex core.\selectlanguage{english}

\subsection*{Coreness disruption predicts cognitive and memory
deficits}\label{coreness-disruption-predicts-cognitive-and-memory-deficits}

\begin{figure*}[h!]
\begin{center}
\includegraphics[width=\textwidth]{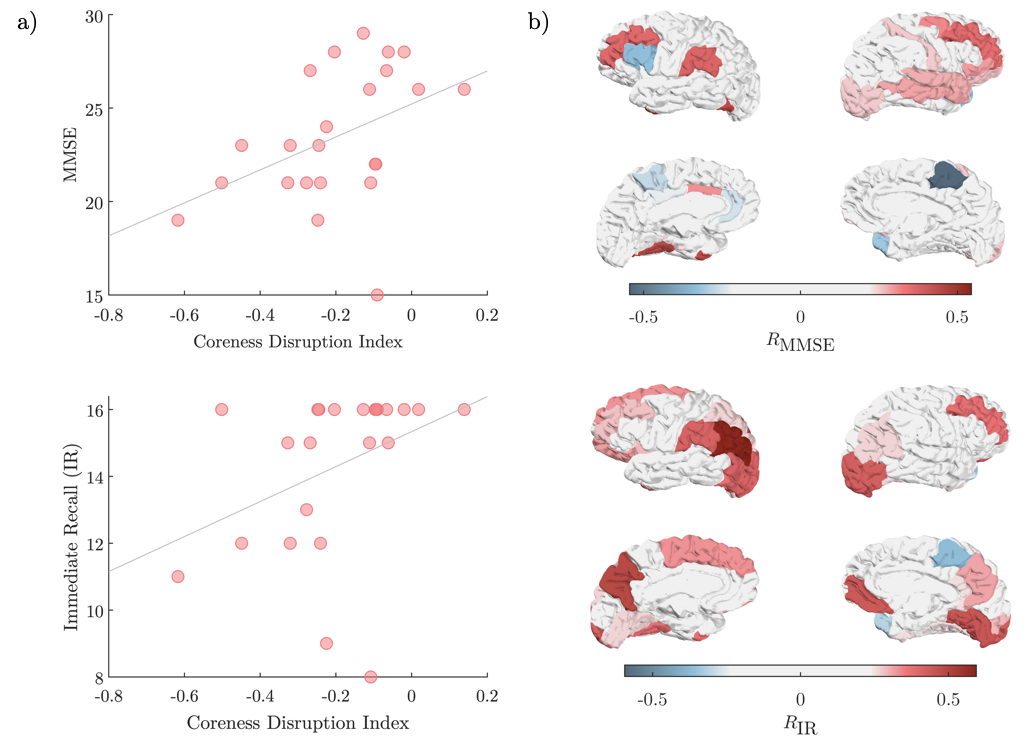}
\caption{{\textbf{Correlation between coreness and cognitive/memory
deficit.}~Panel~\textbf{a)~}show\textbf{~}the values of the~mini-mental
state examination (MMSE) and immediate recall (IR) as a function of
the~coreness disruption index~\(\kappa\).~In panel~\textbf{b)}~
the Spearman correlation values (R) between the regional coreness
\(\mathcal{C}_i\) and the MMSE and IR values are shown over the Desikan
cortical atlas.
{\label{609234}}%
}}
\end{center}
\end{figure*}

We finally conducted a correlation analysis to better understand how the
observed multiplex brain network changes were associated with the
behavioral performance of AD patients. Results show that both cognitive
and memory deficits could be predicted by the individual loss of
regional coreness. At the global scale, the coreness disruption index
significantly correlated with the MMSE (\(R=0.46,p=0.028\)) as well as
with the Immediate (\(R=0.47,p=0.024\)) and Free Recall
(\(R=0.59,p=0.005\)) scores. The higher the \(\kappa\) values,
the better was the performance of the patients (Figure \ref{609234}a,
Supplementary material). At the local scale, temporal, parietal and cortices
were highly positively correlated with the behavior of patients.
Notably, these ROIs overlapped with those exhibiting significant
decreases of regional coreness with respect to healthy controls (Figure
\ref{482763}b). We found similar positive correlations for bilateral
middle frontal ROIs (\(R=0.36,p=0.092\) for left, \(R=0.35,p=0.100\) for
right), while areas in the motor-system appeared not to be involved
except for the paracentral lobule that tended to negatively correlate
with the MMSE (\(R=-0.55,p=0.007\)) and Immediate Recall scores
(\(R=-0.36,p=0.089\)) scores (Figure \ref{609234}b).\selectlanguage{english}

\section*{Discussion}\label{discussion}

\subsection*{Multiplex brain networks}\label{multiplex-brain-networks}

The increasing availability of multimodal neuroimaging data holds a
great potential to enrich our knowledge about fundemental neural
mechanisms and to improve the precision of predictive biomarkers of
brain diseases \cite{Calhoun_2016}. However, how to integrate information
from different neuroimaging modalities is still an open issue. Existing
approaches have mainly focused on merging information at the level of the
native data structure (e.g., signal or images) \cite{Uluda__2014,Biessmann_2011}. Only
recently, investigators have started to propose fusion algorithms in an
effort to infer brain connectivity \cite{Ng_2012} or to detect
mental states \cite{Lei_2011}. Here, we adopted a complementary
solution - based on the nascent field of multilayer network theory - which
preserves the original nature of the different connectivity types.
Similar approaches have been already used in the case of temporal
\cite{Betzel_2017,De_Domenico_2017}, multifrequency \cite{Guillon_2017,de_domenico_mapping_2016} and DTI-fMRI brain
networks \cite{Battiston_2017}. This study considers for the first time
brain networks obtained from three different neuroimaging modalities -
DWI, fMRI and MEG - to construct multiplex brain networks consisting of
nine connectivity layers and to derive an augmented description of their
core-periphery structure in healthy and Alzheimer's diseased subjects. A
crucial step in the multiplex construction is how to weight the
different layers, which typically contain connectivity measured in units (e.g, number of fiber tracks and signal correlation) of different scales. While this is in
general an arbitrary choice, here we established an objective way to
associate a weight to each layer by maximizing the difference of
regional coreness - i.e.~the likelihood of each region to be in the core
- between the groups. Results showed that all the three modalities are
necessary to the group separation. In particular, for MEG only alpha1
was determinant while the other frequency layers had very low, or null,
weights. This is in line with current evidence showing that alpha1
frequency band contains the most discriminant power and connectivity
changes in AD \cite{Babiloni_2004,Blinowska_2017}. Notably, DWI had a very high
contribution coefficient as compared to the other layers. Core-periphery
structure of diffusion-based networks is known to be very robust
\cite{van_den_Heuvel_2011,Hagmann_2008} with respect to functional layers and this might
possibly depend on the heterogeneity of the node degree distribution.
Further resarch is needed to better understand how to normalize connectivity weights between layers when construction multiplex network construction.

\subsection*{Network reorganization in Alzheimer's
disease}\label{network-reorganization-in-alzheimers-disease}

AD is associated with network changes affecting the structure and
function of the brain at multiple spatial and temporal scales
\cite{Stam_2014}. It has been hypothesized that these network
reconfigurations could result from dysconnection patterns initiated by
the gross atrophy of the brain. While several studies have found
significant changes in terms of network efficiency, modularity and node
centrality, the direction of these alterations - in terms of increments
or decrements with respect to healyhy controls - is often unclear and
modality dependent \cite{Tijms_2013}. Here, we focused on the
core-periphery structure of the human brain which has been shown
to have a significant impact on cognition ensuring global integration
across remote cortical areas \cite{van_den_Heuvel_2011}. Structural connectome
studies have reported that AD patients, from the preclinical to dementia
stages, have significant hub-concentrated lesion distributions
\cite{Crossley_2014,Buckner_2009,Dai_2014,Brier_2014,Shu_2018}. However, recent evidence is suggesting that network
disruption is prevalent in the peripheral network components in both AD
\cite{Daianu_2015} and mild cognitive impairement (MCI) patients
\cite{Zhao_2017}. These inconsistent findings suggest that the
network disruption mechanisms remain unclear. By integrating information
from structural and functional brain networks we aimed to overcome this
controversy and provide a more comprehensive insight. Our multiplex
network approach shows that core regions were globally affected in AD
patients as compared to HC subjects and that this result could be
modeled by a global random rewiring process. Specifically, we reported
significant decrements of coreness in temporal and parietal cortices,
which are heavily affected by atrophy processes and beta-amyloid
deposition \cite{Buckner_2008}. However, this change was paralled by a
significant increase of coreness in the paracentral lobules, which originally
belonged to the multiplex periphery. Because regions of the sensorimotor system -
such as paracentral lobule - are not directly affected by the atrophy
process \cite{Agosta2010}, we speculate that possible
compensatory mechanisms could have therefore taken place. In line with
this hypothesis, recent findings suggest that more efficient motor
commands in mild cognitive impaired patients could trigger the later
functional decline \cite{Kubicki_2016}. Longitudinal studies involving
healthy subjects converting into AD will be fundamental to confirm or
reject this prediction \cite{Dubois_2016}.

\par\null

\subsection*{Connectivity-based biomarkers of clinical
behavior}

{\label{490891}}

Brain wiring organization is critically associated with human cognition,
behavior as well as with several neurological and psychiatric
disorders~\cite{Stam_2014}. Network indices describing core-periphery~
and rich-club organization in structural brain networks have been
shown to predict cognitive and motor deficits in multiple
sclerosis~\cite{Stellmann_2017}, and Huntington~disease~\cite{Harrington_2015},
as well as communication impairment in
schizophrenia~\cite{van_den_Heuvel_2013}.~More pertinent to this work, rich-club
biomarkers extracted from DTI networks have been shown to correlate with
cognitive and memory deficits in Alzheimer's disease~\cite{Daianu_2015,tijms_alzheimers_2013,stam_graph_2009}.

Here, we showed that the coreness disruption index - quantifying the
global tendency to weaken core-periphery structure in multimodal brain
networks - determined the cognitive and memory performance of our AD
patients. Patients with a stronger core-periphery organization had
better MMSE and FSCRT scores. At the local scale, temporal, parietal as
well as frontal areas tended to positively correlate with patient's
behavior. These association regions have been shown to be implicated in
the prediction of AD cognitive performance \cite{Khachiyants_2012} and more
in general in memory and language~\cite{Squire_1992,Gordon_1995,Pochon_2002}. We also found
negative correlations with the paracentral lobule (especially right), a
region that is typically involved in motor-related tasks but not in
integrative functions.

From a network perspective, the coreness of regions that tended to be in
the multiplex core - such as temporal, parietal and occipital cortices -
were positively correlated with patients' performance, while among the
peripheral areas the paracentral lobule was negatively correlated with
the behavior. This means that in presence of more severe cognitive and
memory deficits the relative decrease of connectivity in core regions
tended to be replaced by periphery components of the brain system. This
result would confirm the existence of an adjusting mechanism, where the
sensorimotor system might be involved in the compensation of
connectivity loss in systems that are directly impacted by Amyloid-beta
plaques and tau neurofibrillary tangles accumulation~\cite{Iaccarino2018}.

\par\null

\subsection*{Methodological
considerations}

{\label{624728}}

The basic algorithm behind the detection of the core-periphery structure
in multiplex networks is purely deterministic \cite{Ma_2015}. This
means that in principle we could not evaluate the statistical relevance
of the identified structure. To overcome this limitation, we adopted a
procedure that consisted in extracting the core-periphery structure from
a series of multiplex networks obtained by filtering the actual brain multiplex network with increasing density thresholds \cite{Battiston_2018}.
This way we could derive a probabilistic measure of coreness by counting how many
times each ROI was assigned to the core across all the possible
thresholds. For the sake of simplicity we filtered each brain multiplex
by retaining the strongest links so that the average node degree of each
layer ranged from \(k=1\) to \(k=N-1\) (Materials
and methods).

After filtering we did not binarize the surviving links so that we
applied the core-periphery algorithm to sparse weighted multiplex
networks. This approach allows us to exploit all the available information
in the multiplex brain networks. At the same time, we remark that
additional care is needed, as it introduces issues related to the different
nature and distributions of the link weights~\cite{Buld__2018}. Here, we
mitigated this problem by using in the core-periphery algorithm the vector
\(c\) of parameters that can weight the contribution of each
layer (Material and methods). Alternative solutions have been recently
proposed taking into account the normalization of the weight across the
layers by means of singular value decomposition~\cite{Mandke_2018}.

We used an optimization algorithm - namely the particle swarm
optimization - to find the best combination of~\(c\)~components
that maximized the difference of the coreness between AD and HC subjects
(Material and methods). This method presents two limitations that are
important to mention here. First, the time complexity
increases~exponentially with the number of
layers~\(\)\(M\) in order to find a stable
solution. We verified that for~\(M>10\) the research
complexity becomes rapidly intractable due to the large space of
parameter combination to explore.
Second, the cost function optimized by the algorithm and used to evaluate how segregated the two groups are (i.e. two sets of coreness vectors) should be carefully chosen as its accuracy is highly impacted by the size of the feature space (here \(N\), the number of ROIs) and the size of the samples (here the size of the cohorts).
More advanced techniques taking into account the possible nonlinear and/or
non-Euclidean nature of the feature space should be considered for very
large networks~ (e.g., support vector machines, Riemannian geometry).

\par\null

\subsection*{Conclusion}

{\label{786908}}

Consistent with our hypothesis, we have shown that AD atrophy process
generates multimodal connectivity changes that can be quantified by a
multilayer network approach. Specifically we have identified that both
core and - to a minor extent - peripheral cortical areas are affected in
AD, and that the direction of the effect was opposite.~Decrease of
coreness in temporal, parietal and occipital areas - forming the rich
core of the human brain - is paralleled by a possible compensatory
increment in cortical regions that are in the sensorimotor system and
that are more peripheral. These cortical network signatures varied over
individuals and were significant predictors of cognitive and memory
deficits.~ Furthermore, we reported a general framework for the
statistical comparison of core-periphery organization in arbitrary
multiplex networks.~Taken together, our results offer new insights into
the crucial role of core-periphery organization in neurodegenerative
diseases.

\par\null

\section*{Material and methods}\label{material-and-methods}

\subsection*{Cohort inclusion}

{\label{904153}}

The study involved 23 Alzheimer's diseased (AD) patients (13 women) and
26 healthy age-matched control (HC) subjects (19 women). All
participants underwent the Mini-Mental State Examination (MMSE) for
global cognition and the Free and Cued Selective Reminding Test (FCSRT)
for verbal episodic memory. Inclusion criteria for all participants
were:~\emph{i)}~age between 50 and 90;~\emph{ii)}~absence of general
evolutive pathology;~\emph{iii)}~no previous history of psychiatric
diseases;~\emph{iv)}~no contraindication to MRI
examination;~\emph{v)}~French as a mother tongue. Specific criteria for
AD patients were:~\emph{i)}~clinical diagnosis of Alzheimer's
disease;~\emph{ii)}~Mini-Mental State Examination (MMSE) score greater
or equal to 18. All subjects gave written informed consent for
participation in the study, which was approved by the local ethics
committee of the Pitie-Salpetriere Hospital. All experiments were
performed in accordance with relevant guidelines and regulation.

\par\null

\subsection*{Data acquisition and
pre-processing}

{\label{577935}}

Magnetic resonance imaging (MRI) acquisitions were obtained using a 3T
system (Siemens Trio, 32-channel system, with a 12-channel head coil).
The MRI examination included: \emph{(i)} 3D T1-weighted volumetric
magnetization-prepared rapid gradient echo (MPRAGE) sequence with the
following parameters: thickness = 1 mm isotropic, repetition time (TR) =
2300 ms, echo time (TE) = 4.18 ms, inversion time (TI) = 900 ms,
acquisition matrix = 256 \selectlanguage{ngerman}× 256; (\emph{ii)~}echo planar imaging (EPI)
sequence with the following parameters: one image~with no diffusion
sensitization (b0 image) and 50 diffusion-weighted images (DWI) at b =
1500 s/mm\textsuperscript{2}~, thickness = 2 mm isotropic,~TR = 13000
ms, TE = 92 ms, flip angle = 90°, acquisition matrix = 128~×
116;~\emph{(iii)~}functional MRI (fMRI) resting-state
sequence~{sensitive to blood oxygenation level-dependent~(BOLD) contrast
with the following parameters:~}200 images, thickness = 3 mm
isotropic,~TR = 2400 ms, TE = 30 ms, flip angle = 90°, acquisition
matrix = 64~× 64. All MR All MR images were processed using the
\href{http://clinica.run}{Clinica}
software~(\url{http://www.clinica.run}). We first used
the~\texttt{t1-freesurfer-cross-sectional}~pipeline to process
T1-weighted images. to process T1-weighted images. This pipeline is a
wrapper of different tools of the FreeSurfer software
(\url{http://surfer.nmr.mgh.harvard.edu/})~~\cite{Fischl_2012}. It
includes segmentation of subcortical structures, extraction of cortical
surfaces, cortical thickness estimation, spatial normalization onto the
FreeSurfer surface template (FsAverage), and parcellation of cortical
regions. Functional MRI images pre-processing have been conducted using
the~\texttt{fmri-preprocessing} pipeline. Slice timing correction, head
motion correction and unwarping have been applied using SPM12 tools
(\href{http://www.fil.ion.ucl.ac.uk/spm}{www.fil.ion.ucl.ac.uk/spm}).
Separately the brain mask has been extracted from the T1 image of each
subject using FreeSurfer. The resulting fMRI images have then been
registered to the brain-masked T1 image of each subject using SPM's
registration tool. Finally, diffusion-weighted images have been
processed using the~\texttt{dwi-preprocessing}~pipeline of Clinica. For
each subject, all raw DWI volumes were rigidly registered (6 degrees of
freedom (dof)) to the reference b0 image (DWI volume with no diffusion
sensitization) to correct for head motion. The diffusion weighting
directions were appropriately updated \cite{Leemans_2009}. An affine
registration (12 dof) was then performed between each DWI volume and the
reference b0 to correct for eddy current distortions. These
registrations were done using the FSL flirt tool
(\href{http://www.fmrib.ox.ac.uk/fsl}{www.fmrib.ox.ac.uk/fsl}). To
correct for echo-planar imaging (EPI) induced susceptibility artifacts,
the field map image was used as proposed by~\cite{Jezzard_1995} with the
FSL prelude/fugue tools. Finally, the DWI volumes were corrected for
nonuniform intensity using the ANTs N4 bias correction
algorithm~\cite{Tustison_2010}. A single multiplicative bias field from
the reference b0 image was estimated, as suggested
in~\cite{Jeurissen_2014}.

The magnetoencephalography (MEG) experimental protocol consisted in a
resting-state with eyes-closed (EC). Subjects seated comfortably in a
dimly lit electromagnetically and acoustically shielded room and were
asked to relax. MEG signals were collected using a whole-head MEG system
with 102 magnetometers and 204 planar gradiometers (Elekta Neuromag
TRIUX MEG system) at a sampling rate of 1000 Hz and on-line low-pass
filtered at 330 Hz. The ground electrode was located on the right
shoulder blade. An electrocardiogram (EKG, Ag/AgCl electrodes) was
placed on the left abdomen for artifacts correction and a vertical
electrooculogram (EOG) was simultaneously recorded. Four small coils
were attached to the participant in order to monitor head position and
to provide co-registration with the anatomical MRI. The physical
landmarks (the nasion, the left and right pre-auricular points) were
digitized using a Polhemus Fastrak digitizer (Polhemus, Colchester,
VT).~We extracted three consecutive clean epochs of approximately 2
minutes each.~

Signal space separation was performed using
\href{http://imaging.mrc-cbu.cam.ac.uk/meg/Maxfilter}{MaxFilter} to remove
external noise. We used in-house software to remove cardiac and ocular
blink artifacts from MEG signals by means of principal component
analysis. We visually inspected the preprocessed MEG signals in order to
remove epochs that still presented spurious contamination. At the end of
the process, we obtained a coherent dataset consisting of three clean
preprocessed epochs per subject. We reconstructed the MEG activity on
the cortical surface by using a source imaging
technique~\cite{He1999,Baillet2001}:~\emph{i)~}We used the previously segmented
T1-weighted images of each~single subject~\cite{Fischl2002,Fischl2004}~ to import
cortical surfaces in the Brainstorm software~\cite{Tadel2011}
where~they were modeled with approximately 20000 equivalent current
dipoles (i.e., the vertices of the cortical meshes).~\emph{ii)~}We
applied the~wMNE~(weighted Minimum Norm Estimate) algorithm with
overlapping~spheres~\cite{Lin2006} to solve the linear inverse
problem. Both~magnetometer and gradiometer, whose position has been
registered on the T1 image using the digitized head points, were used to
localize the activity over the cortical surface.

\subsection*{Construction of brain
networks}

{\label{621327}}

We built, for each modality, one or multiple brain connectivity networks
whose nodes are regions of interests (ROIs) defined by the Desikan
cortical atlas parcellation~\cite{Desikan_2006} (\(N=68\)
regions); and links are weighted by a given connectivity measure
estimated between each pair of nodes resulting in \(68\times68\)
fully~symmetric~adjacency matrices.

{In the case~of MEG, we used the spectral coherence as a connectivity
estimator with the following parameters:~}~window length = 2 s,~window
type = sliding Hanning, overlap = 25\% number of FFT points (NFFT) =
2000 for a frequency resolution of 0.5 Hz~between~2 Hz and~45
Hz~included{.~}

We then averaged the connectivity matrices within the following
characteristic frequency bands~\cite{Stam2002,Babiloni2004}:~\emph{delta} (2--4
Hz),~\emph{theta} (4.5--7.5 Hz),~\emph{alpha1} (8--10.5
Hz),~\emph{alpha2} (11--13 Hz),~\emph{beta1} (13.5--20 Hz),~\emph{beta2}
(20.5--29.5 Hz) and~\emph{gamma} (30--45 Hz). We finally averaged the
connectivity matrices across the three available epochs to obtain a
robust estimate of the individual brain networks.

For~fMRI data, we focused our analysis on the scale 2 wavelet
correlation matrices that represented - with a TR = 2400ms - the
functional connectivity in the frequency interval
0.05--0.10Hz~\cite{Biswal_1995,Cordes2001,Achard_2006}.

For DWI data, we used the~\href{http://clinica.run}{Clinica}~software to
estimate the fiber orientation distributions (FODs) using constrained
spherical deconvolution (CSD) algorithm from MRtrix3~\texttt{dwi2fod}
tool and tractography based on iFOD2 algorithm
from~{MRtrix3~\texttt{tckgen}~tool. The connectome is finally estimated
by counting the number of tracts connecting each pair of nodes according
to the given parcellation file using~MRtrix3~}\texttt{tck2connectome}
tool.~

\subsection*{Network methods and
models}

{\label{303802}}

We constructed multiplex brain networks in each subject by spatially
aligning DWI, fMRI and MEG source reconstructed connectivity networks.
This led to the following multiplex network with \(M=9\)
layers: \[
\mathcal{M} = \big\{ W^{[m]}, \forall m \in \{\textrm{MEG}_{\delta},...,\textrm{MEG}_{\gamma}, \textrm{fMRI}, \textrm{DWI}\} \big\},
\] where \(W^{[m]} = \big\{w_{ij}^{[m]}\big\}\) is the connectivity
matrix containing the weights of the connections between the ROIs
\(i\) and \(j\) in the modality
\(m\).

To extract the coreness of the nodes from the resulting multiplex
networks, we followed the procedure described by \citet{Battiston_2018}.
First, we filtered each layer by preserving the strongest weights for a
broad range of increasing thresholds. Specifically, we considered
density-based thresholds so that each layer had the same average node
degree from \(k=1\) to \(k=N-1\). Then, for each
threshold we computed the core-periphery of the filtered multiplex
network by evaluating \textit{(i)} the multiplex richness \(\mu_i\) of node
\(i\), defined as follows:
\[
\mu_i = \sum_{m=1}^{M}{c^{[m]}s_i^{[m]}},
\]
with \(s_i^{[m]}\) the strength of the node in the
\(m\)-th layer, and \(c^{[m]}\) the components of the vector \(c\) that
modulate the contribution of each modality-specific layer. And \textit{(ii)}, similarly
to the original paper, we decomposed the richness function into two components
based on the links of node \(i\) that are going towards nodes with lower
richness and those towards nodes with higher richness \(s^{[m]}=s^{[m]-}+s^{[m]+}\).
Thus, the multiplex richness of a node towards richer nodes is defined as follows:
\[
\mu_i^{+} = \sum_{m=1}^{M}{c^{[m]}s_i^{[m]+}}.
\]
We finally counted the number of times that each node was in the core across all the explored thresholds
and we normalized by the maximum theoretical value. As a result, we obtained the coreness \(\mathcal{C}_i\) that can be written as follow:
\[
\mathcal{C}_i = \frac{1}{N-1} \sum_{k=1}^{N-1}{\delta^{[k]}_i},
\] where \[
\delta^{[k]}_i =  \begin{cases}
                    1, & \text{if node $i$ is in the core for the average node degree $k$}.\\
                    0, & \text{otherwise}.
                \end{cases}
\]

We generated random multiplex networks by shuffling a given percentage
of links separately in each layer starting from the actual multiplex
brain network of the HC group. This way the weight distribution was
unchanged and only the topology of the network is impacted.
Specifically, for each HC individual, we generated \(n_{\textrm{rand}}\) new
randomized multiplexes. We chose the minimum number of randomizations
necessary to obtain a variance approximatively equal to the one observed
in the HC and AD groups. This number was \(n_{\textrm{rand}}=3\) and gave in
total \(N_{\textrm{RA}}=n_{\textrm{rand}} \times N_{\textrm{HC}}=78\) samples.

\subsection*{Particles swarm optimization and statistical
analysis}\label{particles-swarm-optimization-and-statistical-analysis}

We used the PSO algorithm \cite{Kennedy} under the MATLAB(R)
software with the default parameters. The Fisher's criterion
\(F(c)\) was defined as follow: \[
F(c)= \frac{ \left(\bar I_{\textrm{AD}}(c) - \bar I_{\textrm{HC}}(c)\right)^{2} }
           { s_{\textrm{AD}}^{2} + s_{\textrm{HC}}^{2} },
\] with
\(\bar I_{\textrm{Pop}}(c)\), the average local (i.e.~node level) index, here the
coreness \(\mathcal{C}\), over a population \(\textrm{Pop}\), which
in our case belongs to \(\{\textrm{AD}, \textrm{HC}\}\), and, \[
s_{\textrm{Pop}}^{2} = \sum_{s \in \textrm{Pop}}{ (I_{s}(c) - \bar I_{\textrm{Pop}}(c))^{2} },
\] with
\(s\) a subject belonging the population
\(\textrm{Pop}\).

Since, in our case, \(F(c) = F(ac), \forall a \in \mathbb{R}^{+}\), and in order to save one
dimension in the searching space, we expressed the coefficient
\(c\) as a point on the positive section of the unitary
hypersphere of dimension \(M=9\) such that:
\[
c = \begin{pmatrix}
   \sin \phi_1 ... \sin \phi_{8}  \\
   \sin \phi_1 ... \sin \phi_{7} \cos \phi_{8}  \\
   \sin \phi_1 ... \sin \phi_{6} \cos \phi_{7}  \\
   \sin \phi_1 ... \sin \phi_{5} \cos \phi_{6}  \\
   \sin \phi_1 ... \sin \phi_{4} \cos \phi_{5}  \\
   \sin \phi_1 ... \sin \phi_{3} \cos \phi_{4}  \\
   \sin \phi_1 \sin \phi_{2} \cos \phi_{3}  \\
   \sin \phi_1 \cos \phi_{2}  \\
   \cos \phi_1
\end{pmatrix}, \phi_k \in \Big[0, \frac{\pi}{2}\Big], \forall k \in [1,M-1].
\]

To consider the non-gaussian nature of the data we considered
non-parametric statistics when assessing differences between populations
and prediction of behavioral scores. To these ends, we used respectively
permutation t-tests and Spearman correlation coefficients. The
statistical threholds were set to \(\alpha=0.05\) and we applied a
rough false discovery rate (FDR) correction to account for the
\(N=68\) post-hoc tests at the level of brain regions
(\(\alpha_{FDR}=0.025\)).

\par\null

\pagebreak

\section*{Acknowledgements}

{\label{427711}}

The research leading to these results has received funding from the
program ``Investissements d'avenir'' ANR-10-IAIHU-06 (Agence Nationale
de la Recherche-10-IA Institut Hospitalo-Universitaire-6),
ANR-11-IDEX-004 (Agence Nationale de la Recherche-11- Initiative
d'Excellence-004, project LearnPETMR number SU-16-R-EMR-16)~~and~from
Agence Nationale de la Recherche (project HM-TC, grant number
ANR-09-EMER-006). FD acknowledges support from the ``Agence Nationale de
la Recherche'' through contract number ANR-15-NEUC-0006-02. The content
is solely the responsibility of the authors and does not necessarily
represent the official views of any of the funding agencies.

\par\null\par\null\par\null\par\null

\selectlanguage{english}
\bibliographystyle{plainnat}
\bibliography{ms}

\end{document}


\title{Disrupted core-periphery structure of multimodal brain networks in
Alzheimer's Disease (Supplementary Material)}

\author[1,2]{Jeremy Guillon}%
\author[3]{Mario Chavez}%
\author[10,3,2]{Federico Battiston}%
\author[4]{Yohan Attal}%
\author[5,6,7]{Valentina La Corte}%
\author[1]{Michel Thiebaut de Schotten}%
\author[8]{Bruno Dubois}%
\author[9]{Denis Schwartz}%
\author[2,1]{Olivier Colliot}%
\author[2,1,*]{Fabrizio De Vico Fallani}%
\affil[1]{Institut du Cerveau et de la Moelle Epiniere, ICM, Inserm, U 1127, CNRS, UMR 7225, Sorbonne Universite, F-75013, Paris, France}
\affil[2]{Inria Paris, Aramis project-team, F-75013, Paris, France}
\affil[3]{CNRS, UMR 7225, F-75013, Paris, France}
\affil[4]{MyBrain Technologies, Paris, France}
\affil[5]{Department of Neurology, Institut de la Memoire et de la Maladie d’Alzheimer - IM2A, Paris, France}%
\affil[6]{INSERM UMR 894, Center of Psychiatry and Neurosciences, Memory and Cognition Laboratory, Paris, France}%
\affil[7]{Institute of Psychology, University Paris Descartes, Sorbonne Paris Cite, France}%
\affil[8]{Institut de la Mémoire et de la Maladie d’Alzheimer - IM2A, AP-HP, Sorbonne Université, Paris, France}%
\affil[9]{Institut du Cerveau et de la Moelle Epiniere, ICM, Inserm U 1127, CNRS UMR 7225, Sorbonne Universite, Ecole Normale Superieure, ENS, Centre MEG-EEG, F-75013, Paris, France}
\affil[10]{Department of Network and Data Science, Central European University, Budapest 1051, Hungary}
\affil[*]{Corresponding author}

\vspace{-1em}

  \date{\today}

\begingroup
\let\center\flushleft
\let\endcenter\endflushleft
\maketitle
\endgroup

\pagebreak


\section*{Supplementary material}

{\label{105296}}

\subsection*{Supplementary figure}

{\label{494595}}\par\null\selectlanguage{english}
\begin{figure}[h!]
\begin{center}
\includegraphics[width=1.00\columnwidth]{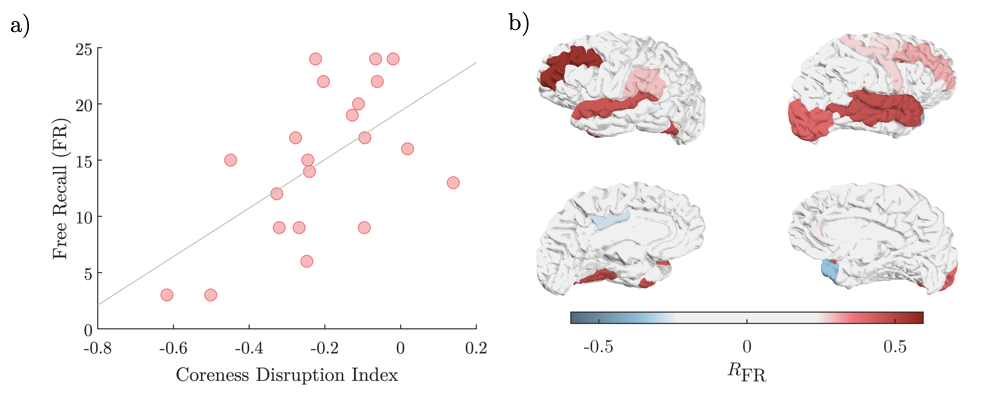}
\caption{{\textbf{Correlation between corenes and cognitive/memory
deficit.~}Panel~\textbf{a)~}shows the values of the free recall (FR) as
a function of the coreness diruption index~\(\kappa\) .~In
panel~\textbf{b)~}the Spearman correlation values (\(R_\textrm{FR}\))
between the regional coreness~\(\mathcal{C}_i\) and the FR values are
shown over the Desikan cortical atlas.
{\label{840418}}{\label{840418}}%
}}
\end{center}
\end{figure}

\par\null\par\null\par\null

\subsection*{Supplementary~tables}

{\label{170658}}\selectlanguage{english}
\begin{table}[]
\caption{{ADs population coreness disruption indices $\kappa$ and the associated $R^2$ of the linear regression.}}
\label{corenessdisruptionindextable}
\begin{tabular}{r|c|c}
Subject ID & $\kappa$ & $R^2$ \\ \hline\hline
1           & -0,095     & 0,018 \\
2           & -0,111     & 0,033 \\
3           & -0,096     & 0,019 \\
4           & 0,139      & 0,040 \\
5           & -0,502     & 0,412 \\
6           & 0,018      & 0,001 \\
7           & -0,327     & 0,218 \\
8           & -0,617     & 0,267 \\
9           & -0,241     & 0,106 \\
10          & -0,062     & 0,015 \\
11          & -0,204     & 0,099 \\
12          & -0,245     & 0,085 \\
13          & -0,225     & 0,056 \\
14          & -0,020     & 0,001 \\
15          & -0,128     & 0,019 \\
16          & -0,248     & 0,141 \\
17          & -0,268     & 0,106 \\
18          & -0,278     & 0,081 \\
19          & -0,449     & 0,387 \\
20          & -0,091     & 0,011 \\
21          & -0,108     & 0,025 \\
22          & -0,320     & 0,258 \\
23          & -0,066     & 0,014
\end{tabular}
\end{table}\selectlanguage{english}

\begin{table}[]
\caption{{Average coreness differences and their associated $p$-value. Ordered in descending $\bar{\mathcal{C}}_{\textrm{HC},i}$. Significant values ($\alpha_\textrm{FDR}=0,025$) are in bold.}}
\label{corenessdifftable}
\scriptsize
\begin{tabular}{r|c|c|c|c}
Label              & $\bar{\mathcal{C}}_{\textrm{AD},i}$ & $\bar{\mathcal{C}}_{\textrm{HC},i}$ & $\bar{\mathcal{C}}_{\textrm{AD},i}-\bar{\mathcal{C}}_{\textrm{HC},i}$ & $p$-Value \\ \hline\hline
superiorfrontal R          & 0,923          & 0,952          & -0,029          & 0,743  \\
precentral R               & 0,883          & 0,921          & -0,038          & 0,719  \\
superiorfrontal L          & 0,846          & 0,921          & -0,075          & 0,579  \\
precentral L               & 0,695          & 0,916          & -0,221          & 0,543  \\
\textbf{superiortemporal R}         & 0,609          & 0,862          & -0,253          & \textbf{0,005}  \\
middletemporal R           & 0,677          & 0,855          & -0,178          & 0,121  \\
superiorparietal R         & 0,781          & 0,806          & -0,025          & 0,650  \\
\textbf{lateraloccipital L}         & 0,562          & 0,794          & -0,233          & \textbf{0,020}  \\
postcentral R              & 0,696          & 0,778          & -0,082          & 0,613  \\
\textbf{lateraloccipital R}         & 0,520          & 0,776          & -0,256          & \textbf{0,008}  \\
inferiorparietal R         & 0,535          & 0,763          & -0,228          & 0,071  \\
fusiform L                 & 0,575          & 0,720          & -0,145          & 0,133  \\
lingual L                  & 0,558          & 0,715          & -0,157          & 0,210  \\
superiortemporal L         & 0,577          & 0,703          & -0,126          & 0,316  \\
\textbf{middletemporal L}           & 0,411          & 0,681          & -0,270          & \textbf{0,017}  \\
lingual R                  & 0,445          & 0,660          & -0,214          & 0,034  \\
fusiform R                 & 0,494          & 0,650          & -0,156          & 0,203  \\
\textbf{inferiortemporal R}         & 0,344          & 0,641          & -0,297          & \textbf{0,002}  \\
superiorparietal L         & 0,632          & 0,637          & -0,005          & 0,992  \\
precuneus R                & 0,598          & 0,631          & -0,033          & 0,756  \\
inferiortemporal L         & 0,478          & 0,569          & -0,091          & 0,288  \\
supramarginal R            & 0,482          & 0,513          & -0,032          & 0,674  \\
inferiorparietal L         & 0,401          & 0,502          & -0,101          & 0,279  \\
postcentral L              & 0,493          & 0,501          & -0,009          & 0,787  \\
precuneus L                & 0,296          & 0,437          & -0,142          & 0,173  \\
insula R                   & 0,404          & 0,398          & 0,006           & 0,666  \\
\textbf{pericalcarine L}            & 0,126          & 0,379          & -0,254          & \textbf{0,012}  \\
rostralmiddlefrontal R     & 0,342          & 0,334          & 0,008           & 0,880  \\
pericalcarine R            & 0,176          & 0,304          & -0,128          & 0,151  \\
insula L                   & 0,369          & 0,290          & 0,079           & 0,388  \\
supramarginal L            & 0,255          & 0,270          & -0,015          & 0,771  \\
lateralorbitofrontal L     & 0,223          & 0,254          & -0,031          & 0,747  \\
rostralmiddlefrontal L     & 0,299          & 0,213          & 0,086           & 0,349  \\
cuneus R                   & 0,091          & 0,191          & -0,099          & 0,391  \\
isthmuscingulate L         & 0,134          & 0,186          & -0,052          & 0,324  \\
posteriorcingulate L       & 0,116          & 0,172          & -0,056          & 0,559  \\
medialorbitofrontal L      & 0,226          & 0,158          & 0,067           & 0,221  \\
medialorbitofrontal R      & 0,178          & 0,136          & 0,042           & 0,709  \\
lateralorbitofrontal R     & 0,109          & 0,131          & -0,022          & 0,696  \\
isthmuscingulate R         & 0,086          & 0,122          & -0,037          & 0,907  \\
cuneus L                   & 0,100          & 0,118          & -0,018          & 0,201  \\
posteriorcingulate R       & 0,174          & 0,115          & 0,059           & 0,192  \\
caudalmiddlefrontal R      & 0,040          & 0,086          & -0,045          & 0,966  \\
\textbf{paracentral R}              & 0,151          & 0,076          & 0,074           & \textbf{0,022}  \\
parsopercularis R          & 0,010          & 0,057          & -0,048          & 0,337  \\
parsopercularis L          & 0,049          & 0,046          & 0,002           & 1,000  \\
caudalmiddlefrontal L      & 0,047          & 0,041          & 0,006           & 0,946  \\
bankssts R                 & 0,018          & 0,033          & -0,015          & 0,326  \\
rostralanteriorcingulate L & 0,032          & 0,031          & 0,001           & 0,568  \\
bankssts L                 & 0,010          & 0,021          & -0,012          & 0,588  \\
parahippocampal L          & 0,025          & 0,021          & 0,005           & 0,928  \\
parstriangularis R         & 0,002          & 0,015          & -0,014          & 0,754  \\
paracentral L              & 0,036          & 0,015          & 0,021           & 0,462  \\
parahippocampal R          & 0,035          & 0,014          & 0,021           & 0,679  \\
parstriangularis L         & 0,007          & 0,010          & -0,003          & 0,909  \\
caudalanteriorcingulate L  & 0,014          & 0,009          & 0,004           & 0,629  \\
caudalanteriorcingulate R  & 0,067          & 0,008          & 0,059           & 0,929  \\
temporalpole L             & 0,008          & 0,007          & 0,001           & 0,805  \\
rostralanteriorcingulate R & 0,021          & 0,005          & 0,016           & 0,264  \\
entorhinal L               & 0,004          & 0,005          & -0,001          & 0,399  \\
transversetemporal L       & 0,000          & 0,004          & -0,004          & 0,189  \\
frontalpole L              & 0,000          & 0,003          & -0,003          & 0,368  \\
parsorbitalis L            & 0,004          & 0,003          & 0,001           & 0,874  \\
temporalpole R             & 0,010          & 0,002          & 0,008           & 0,682  \\
parsorbitalis R            & 0,001          & 0,002          & -0,002          & 0,630  \\
transversetemporal R       & 0,001          & 0,002          & -0,002          & 0,630  \\
entorhinal R               & 0,000          & 0,000          & 0,000           & NA    \\
frontalpole R              & 0,000          & 0,000          & 0,000           & NA
\end{tabular}
\end{table}

\selectlanguage{english}
\clearpage